\documentclass[aps,prl,twocolumn,superscriptaddress,longbibliography]{revtex4-1}
\usepackage{graphicx}
\usepackage{amssymb,amsmath}
\usepackage{bm}
\usepackage{dcolumn,dsfont}
\usepackage{subfigure}
\usepackage{float}
\usepackage[OT1]{fontenc} 
\usepackage{url}
\usepackage{mathrsfs}
\usepackage{slashed}
\usepackage{color}
\usepackage{verbatim}
\usepackage{enumitem}
\usepackage{txfonts}
\usepackage{soul,physics}
\usepackage[driverfallback=dvipdfm]{hyperref}
\hypersetup{pdfpagemode=FullScreen,colorlinks=true,breaklinks,urlcolor=blue,linkcolor=blue,citecolor=blue}
\usepackage{natbib}
\usepackage{environ}

\usepackage{subfigure}
\usepackage{comment}
\usepackage{hyperref}
\usepackage{amssymb}
\usepackage{bm}
\usepackage{graphicx}
\usepackage{amsmath,amssymb}
\usepackage{tikz,fp}
\usepackage{tikz-cd}
\usetikzlibrary{arrows}
\usetikzlibrary{intersections}
\usetikzlibrary{shapes.geometric}
\usetikzlibrary{decorations.pathmorphing, patterns,shapes,fixedpointarithmetic}
\usetikzlibrary{decorations.markings}

\pgfdeclarepatternformonly{south west lines}{\pgfqpoint{-0pt}{-0pt}}{\pgfqpoint{3pt}{3pt}}{\pgfqpoint{3pt}{3pt}}{
    \pgfsetlinewidth{0.4pt}
    \pgfpathmoveto{\pgfqpoint{0pt}{0pt}}
    \pgfpathlineto{\pgfqpoint{3pt}{3pt}}
    \pgfpathmoveto{\pgfqpoint{2.8pt}{-.2pt}}
    \pgfpathlineto{\pgfqpoint{3.2pt}{.2pt}}
    \pgfpathmoveto{\pgfqpoint{-.2pt}{2.8pt}}
    \pgfpathlineto{\pgfqpoint{.2pt}{3.2pt}}
    \pgfusepath{stroke}}

\tikzset{
    mid arrow/.style={postaction={decorate,decoration={
                markings,
                mark=at position .575 with {\arrow{stealth}}
    }}},
    near arrow/.style={postaction={decorate,decoration={
                markings,
                mark=at position .275 with {\arrow{stealth}}
    }}},
    far arrow/.style={postaction={decorate,decoration={
                markings,
                mark=at position .800 with {\arrow{stealth}}
    }}},
    snake arrow/.style={fixed point arithmetic, decorate, decoration={snake,amplitude=2pt, segment length=11pt},postaction={decoration={markings,mark=at position 0.625 with {\arrow{stealth}}},decorate}},
}
\tikzset{
  baseline = -0.5ex,
  wavy/.style = {
    thick,
    decorate,
    decoration={snake,amplitude=2pt,segment length=5pt}},
  sdot/.style = {
    circle,
    draw=none,
    fill=black,
    minimum size=2.5pt,
    inner sep=0pt},
  bdot/.style = {
    circle,
    draw=none,
    fill=black,
    minimum size=4pt,
    inner sep=0pt},
  svertex/.style = {
    circle,
    draw=black,
    thick,
    fill=lightgray,
    minimum size=14pt,
    inner sep=1pt},
  bvertex/.style = {
    circle,
    draw=black,
    thick,
    fill=lightgray,
    minimum size=24pt},
  bvertexsmall/.style = {
    circle,
    draw=black,
    thick,
    fill=lightgray,
    minimum size=7pt},
  bvertexnormal/.style = {
    circle,
    draw=black,
    thick,
    fill=lightgray,
    minimum size=16pt},
    bvertexnormal2/.style = {
    circle,
    draw=black,
    thick,
    fill=lightgray,
    minimum size=24pt},
  dvertex/.style = {
    circle,
    draw=black,
    thick,
    fill=gray,
    minimum size=25pt}}

\NewEnviron{myequation}{%
    \begin{equation}
    \scalebox{0.91}{$\displaystyle{\BODY}$}
    \end{equation}
    }

\begin{document}
    
    \title{Distinguishing Quantum Phases through Cusps in Full Counting Statistics }
 
    \author{Chang-Yan Wang}
    \thanks{They contribute equally to this work.}
    \affiliation{Institute for Advanced Study, Tsinghua University, Beijing, 100084, China}
    
    \author{Tian-Gang Zhou}
    \thanks{They contribute equally to this work.}
    \affiliation{Institute for Advanced Study, Tsinghua University, Beijing, 100084, China}
    
    \author{Yi-Neng Zhou}
    \thanks{They contribute equally to this work.}
    \affiliation{Institute for Advanced Study, Tsinghua University, Beijing, 100084, China}
    
        \author{Pengfei Zhang}
    \thanks{pengfeizhang.physics@gmail.com}
    \affiliation{Department of Physics, Fudan University, Shanghai, 200438, China}
    \affiliation{Center for Field Theory and Particle Physics, Fudan University, Shanghai, 200438, China}
    \affiliation{Shanghai Qi Zhi Institute, AI Tower, Xuhui District, Shanghai 200232, China}
    
    \date{\today}
    \begin{abstract}
    Measuring physical observables requires averaging experimental outcomes over numerous identical measurements. The complete distribution function of possible outcomes or its Fourier transform, known as the full counting statistics, provides a more detailed description. This method captures the fundamental quantum fluctuations in many-body systems and has gained significant attention in quantum transport research. In this letter, we propose that cusp singularities in the full counting statistics are a novel tool for distinguishing between ordered and disordered phases. As a specific example, we focus on the superfluid-to-Mott transition in the Bose-Hubbard model and introduce $\mathcal{Z}_A(\alpha)=\langle \exp({i\alpha \sum_{i\in A}(\hat{n}_i}-\overline{n}))\rangle $ with $\overline{n}=\langle n_i \rangle$. Through both analytical analysis and numerical simulations, we demonstrate that $\partial_\alpha \log \mathcal{Z}_A(\alpha)$ exhibits a discontinuity near $\alpha=\pi$ in the superfluid phase when the subsystem size is sufficiently large, while it remains smooth in the Mott phase. This discontinuity can be interpreted as a first-order transition between different semi-classical configurations of vortices. We anticipate that our discoveries can be readily tested using state-of-the-art ultracold atom and superconducting qubit platforms.
    \end{abstract}
    
    \maketitle

    \emph{ \color{blue!60}Introduction.--}
    Fluctuations are pervasive in quantum many-body systems and serve as a window into fundamental physical principles. For example, in quasi-one-dimensional electronic systems, charge transfer exhibits non-trivial fluctuations around its expected value. This phenomenon, indicative of charge quantization and known as shot noise, has been extensively studied \cite{kamenev2023field}. Full counting statistics (FCS), a theoretical framework involving the Fourier transform of the charge distribution, has been introduced as a comprehensive method for describing these complex charge fluctuations \cite{belzig_full_2001, cherng_quantum_2007, kitagawa_dynamics_2011, levitov_counting_2004, shelankov_charge_2003, barratt_field_2022, bertini_nonequilibrium_2023, calabrese_exact_2012, eisler_full_2013, eisler_universality_2013, ivanov_characterizing_2013, klich_note_2014, levine_full_2012, levitov_electron_1996, mcculloch_full_2023, oshima_charge_2023, klich_quantum_2009, song_bipartite_2012, song_entanglement_2011, song_general_2010, susstrunk_free_2013}. FCS is defined as follows:
    \begin{equation}
     \mathcal{Z}_A(\alpha)=\langle e^{i\alpha \sum_{i\in A}(\hat{n}_i-\overline{n})}\rangle\equiv e^{-\mathcal{F}_A(\alpha)},
    \end{equation}
    where $\alpha \in (-\pi,\pi]$ and $\overline{n}=\langle n_i \rangle$ is the filling fraction of the system. Up to a numerical factor, $\mathcal{Z}_A$ is a Fourier transform of $p_A(n)$, the probability of finding $n$ charges in the subsystem $A$. Therefore, it provides complete information about the distribution of charges. 
    
\begin{figure}
    \centering
    \includegraphics[width=0.95\linewidth]{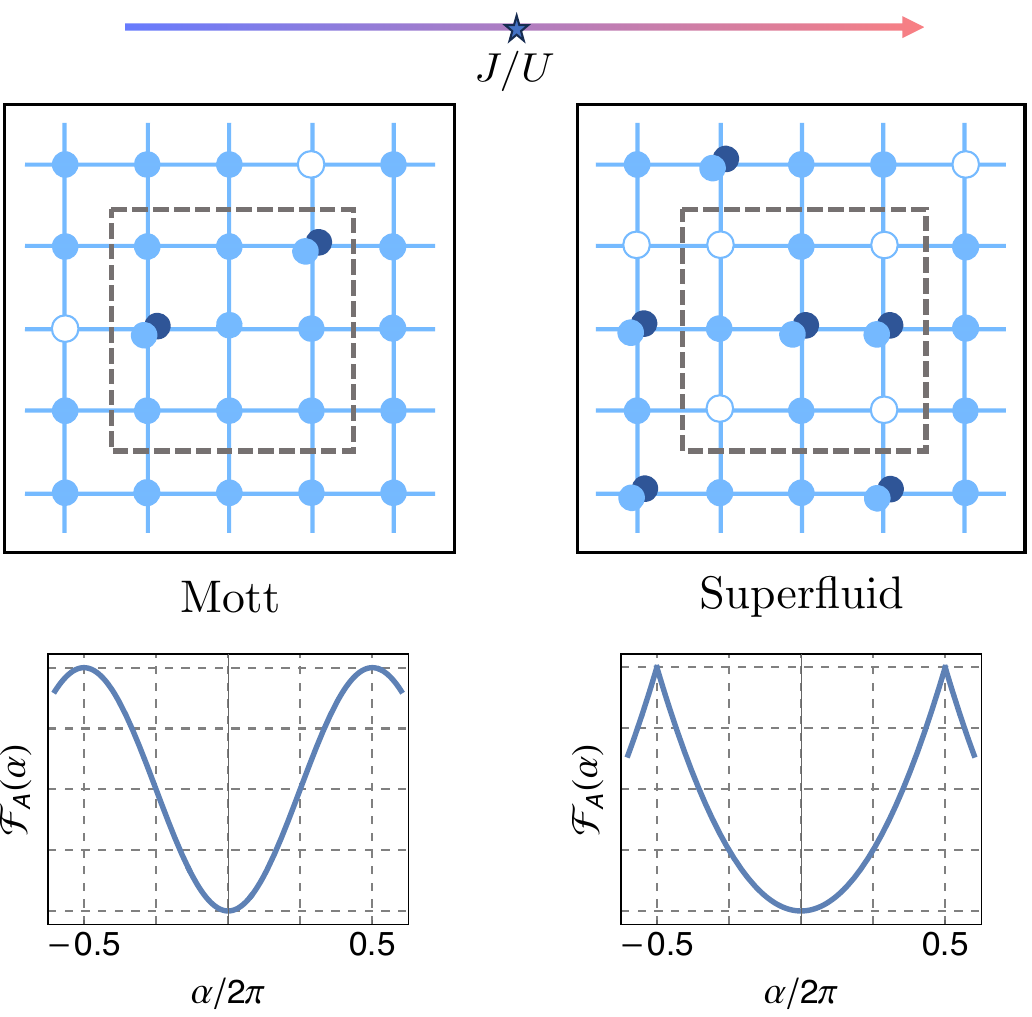}
    \caption{The schematics depict the FCS in both the superfluid and Mott phases of the Bose-Hubbard model. The area marked by the dashed line is subsystem A. In the superfluid phase, the FCS displays a cusp near $\alpha=\pm \pi$, which serves as a non-local order parameter. In the Mott phase, the FCS is instead an analytical function of $\alpha$. } \label{fig1}
\end{figure}

    Although the FCS was originally introduced for studying charge transport, it has gained crucial importance in modern condensed matter theory, often referred to as disorder operators \cite{fradkin_disorder_2017}. The dependence of the FCS on subsystem size has been systematically explored to characterize quantum phase \cite{wang_scaling_2021, zhao_higherform_2021, wu_universal_2021, wu_categorical_2021, liu_disorder_2023,liu_fermion_2023, wang_scaling_2022} and gapped topological phases \cite{chen_topological_2022}. Nevertheless, the primary focus is directed towards the limit of small $\alpha$. Recent studies have expanded the application of FCS to explore measurement-induced phase transitions \cite{tirrito_full_2023,Tian-GangZhou:2023jsj}, continuing to establish its connection with entanglement entropy in non-interacting systems. \cite{klich_quantum_2009,song_entanglement_2011,song_bipartite_2012}.
    
    In this study, we explore the $\alpha$ dependence of the FCS in models that undergo order-disorder quantum phase transitions, focusing on systems possessing U(1) symmetry. To reveal universal features, we concentrate on the $\alpha$ dependency near $\alpha=\pm \pi$. We propose that the FCS displays non-analytic behavior in the ordered phase while maintaining a smooth function in the disordered phase. In particular, the discontinuity of the first order derivative for sufficiently large subsystem size $L_A$
    \begin{equation}\label{eq:Delta}
        \Delta\equiv \lim_{\epsilon\rightarrow 0}\lim_{L_A\rightarrow \infty}\Big[\partial_\alpha \mathcal{F}_A(\pi-\epsilon)-\partial_\alpha \mathcal{F}_A(-\pi+\epsilon)\Big],
    \end{equation}
    can serve as a non-local order parameter for order-disorder transitions, regardless of dimensionality, as depicted in FIG \ref{fig1}. This is intuitive since phases with and without order exhibit different amounts of fluctuation, making FCS an apt probe for distinguishing quantum phases. We illustrate our proposition using the example of the Bose-Hubbard model, combining analytical analysis and numerical simulations. As we will elucidate, the analyticity arises from the first-order transition in the configuration of vortices in the superfluid phase as $\alpha$ varies across $\alpha=\pm \pi$, analogous to the Page curve as a function of subsystem size \cite{page_information_1993, page_time_2013}. Our theoretical proposal can be readily tested in ultracold atom experiments.
    
    
    \emph{ \color{blue!60}Model.--}
    A prominent example for order-disorder transition is the superfluid-to-Mott transition for bosons in optical lattices, described by the Bose-Hubbard model \cite{fisher_boson_1989}:
    \begin{equation}
  \hat{H} = -J \sum_{\langle ij\rangle} (\hat{b}_i^\dag \hat{b}_j + \mathrm{h.c.}) + \frac{U}{2}\sum_i \hat{n}_i (\hat{n}_i - 1).
    \end{equation}
    We focus on the integer filling case. The system enters a superfluid phase when $J/U>r_c$, where charge fluctuations are coupled to a gapless phonon mode. In 1D, the superfluid phase processes quasi-long-range order, and previous studies suggest $r_c\approx 0.28$ \cite{PhysRevB.58.R14741, cazalilla_one_2011, batrouni_quantum_1990, batrouni_worldline_1992, elstner_dynamics_1999, kashurnikov_mottinsulatorsuperfluidliquid_1996, kashurnikov_exact_1996, kuhner_onedimensional_2000}. In higher dimensions, the ground state experiences a spontaneous symmetry breaking, characterized by a non-vanishing order parameter. On the contrary, the system is in a Mott phase for $J/U<r_c$, which preserves the U(1) symmetry and displays a finite charge gap. 
    The model has been realized in ultracold atoms \cite{jaksch_cold_1998, bakr_probing_2010, greiner_quantum_2002, bloch_ultracold_2005, sherson_singleatomresolved_2010, weitenberg_singlespin_2011, endres_observation_2011, stoferle_transition_2004,PhysRevX.13.021042} and superconducting qubits \cite{fedorov_photon_2021, yanay_twodimensional_2020, ye_propagation_2019}. Notably, a recent experiment \cite{PhysRevX.13.021042} probes the phase transition in 2D by measuring the brane parity order $P_A= \langle \exp({i\pi \sum_{i\in A}\hat{n}_i})\rangle $, focusing on scaling with $L_A$ for a subsystem $A$ containing $L_A^D$ sites \cite{RATH2013256,PhysRevLett.118.157602,PhysRevB.94.085119}. Similar scaling behavior has been reported for the entanglement entropy of Fermi liquids \cite{PhysRevLett.96.010404,PhysRevLett.96.100503,PhysRevLett.105.050502} and the steady states of free fermions under non-unitary dynamics \cite{Zhang:2021klq,PhysRevResearch.2.033017,PhysRevLett.126.170602}.
    
     \begin{figure}[t]
    \centering
    \includegraphics[width=0.98\linewidth]{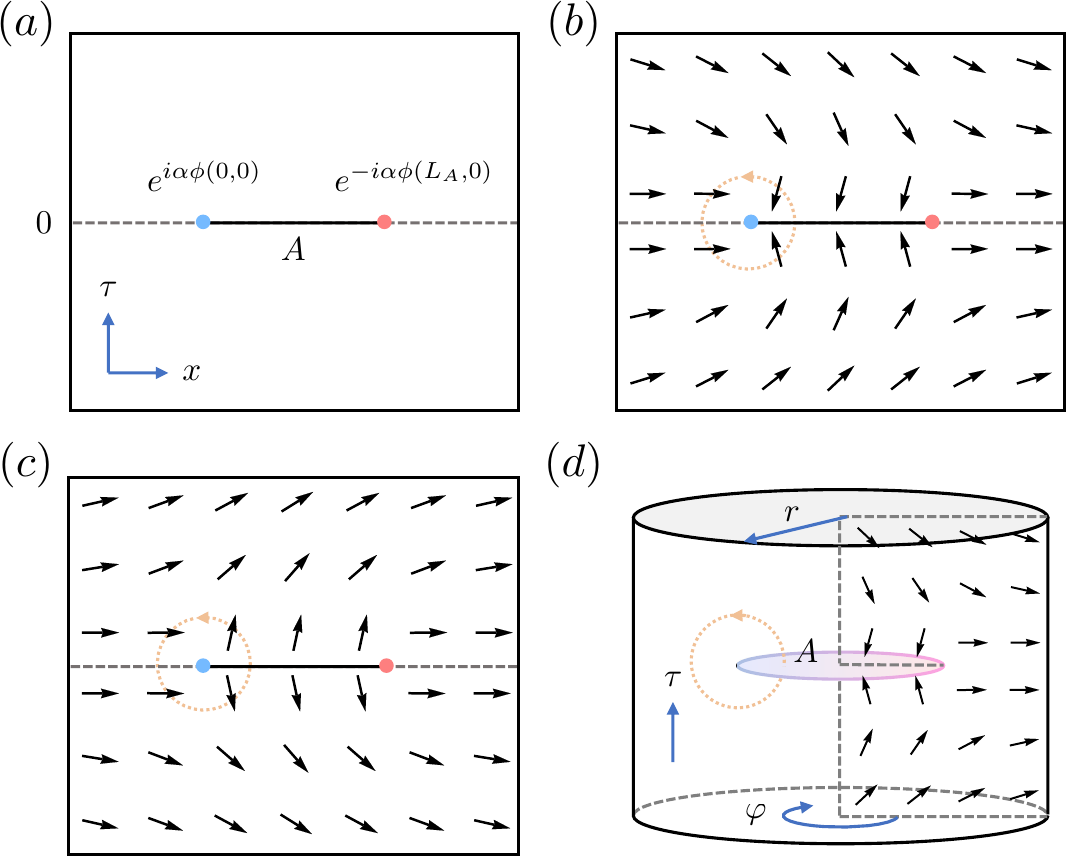}
    \caption{ Three different pictures are considered for the calculation of the FCS: (a) a two-point function on the infinite plane, (b-c) two different semi-classical configurations of $\theta(x,\tau)$ that dominate the FCS near $\alpha=\pm \pi$. (d) semi-classical configurations of $\theta(x,\tau)$ in 2D with a disk-like subregion $A$. In (b-d), the winding number of the vortices or the vortex loop can be measured along the yellow loop.
 } \label{fig2}
\end{figure}
    \vspace{5pt}

    \emph{ \color{blue!60}Superfluid phase.--}
    We first study the FCS in the superfluid phase. Since the dominant contribution comes from the phonon mode, we adopt the field theory description with an effective action \cite{altland_condensed_2023}
    \begin{equation}\label{eq:eff}
    S_{\text{eff}}=\frac{\rho_s}{2}[(\partial_\tau \theta)^2/u^2+(\bm{\nabla} \theta)^2].
    \end{equation}
    Here, we employ the imaginary-time path integral approach. $\theta(\bm{x},t) \in (-\pi,\pi]$ is the field for phase fluctuation of the superfluid. $\rho_s$ is the superfluid density and $u$ is the phonon velocity. Let us first focus on the 1D case, where we can identify the Luttinger parameter $K=\pi \rho_s/u$ \cite{giamarchi_quantum_2004, cazalilla_one_2011}. The generalization to higher dimensions will be discussed subsequently.

    We demonstrate the non-analyticity of $\mathcal{F}_A(\alpha)$ in 1D by providing two complimentary pictures, as shown in FIG \ref{fig2} (a-c). We begin with a straightforward calculation of the FCS using the Luttinger liquid theory. In the continuum limit, the density field can be approximated as $n(x)=\overline{n}-\frac{1}{\pi}\nabla \phi(x)$ \cite{giamarchi_quantum_2004, cazalilla_one_2011}. Here, $\phi(x)$ is the dual field of $\theta(x)$, which satisfies the commutation relation $[\phi(x),\nabla \theta(y)]=i\pi\delta(x-y)$. Therefore, the FCS can be expressed as $\mathcal{Z}_A^{\text{LL}}(\alpha)=\langle e^{i\alpha [\phi(0,0)-\phi(L_A,0)]}\rangle_p$,
    where we prepare the ground state by employing an imaginary-time path integral over a half-infinite plane and then insert the charge operator at $\tau=0$. An illustration is provided in FIG \ref{fig2} (a). For the quadratic action given by Eq. \eqref{eq:eff}, a direction calculation yields \cite{giamarchi_quantum_2004, cazalilla_one_2011}
    \begin{equation}\label{eq:Fquadratic}
    \mathcal{F}^{\text{LL}}_A(\alpha)=\frac{K\alpha^2}{4}\ln\bigg(\frac{L_A^2+a^2}{a^2}\bigg).
    \end{equation}
    Here, $a$ serves as a short-distance cutoff introduced for regularization. Recalling that $\alpha\in(-\pi,\pi]$, this result predicts a cusp near $\alpha=\pm \pi$, which gives $\Delta= 2\pi K \ln L_A$. A similar phenomenon has been observed in the volume-law entangled phase of non-Hermitian Hamiltonians \cite{Tian-GangZhou:2023jsj}. Previous studies on 1D fermionic models \cite{Jiang:2022tmb} also unveils a quadratic dependence of $\alpha$ using the Widom-Sobolev formula or bosonization, although main attention has been paid to small $\alpha$. 

     \vspace{5pt}
    \emph{ \color{blue!60}Finite-$L_A$ corrections.--}
    In the above calculation, the periodicity of $F(\alpha)$ is enforced by hand, which can raise concerns about whether the non-analyticity at $\alpha=\pi$ is an artifact of the field theory calculation. Furthermore, there is a general belief that all physical observables should exhibit smooth behavior for finite system sizes, especially in lattice systems. In other words, there should be corrections accounting for finite $L_A/a$. Last but not least, a direct inverse Fourier transform shows that Eq. \eqref{eq:Fquadratic} predicts a negative probability for finding a large number of charges in subsystem A, which is unphysical. To address these questions, we study the finite-size correction of the FCS. The result further provides a semi-classical understanding of the singularity's presence.

    To begin with, we express the insertion of $\exp({i\pi \sum_{i\in A}\hat{n}_i})$ as a change in boundary conditions for the phase field $\theta(x,\tau)$:
    \begin{equation}\label{eq:bdy}
    \theta(x,0^+)=\theta(x,0^-)+\alpha~\Theta(x)\Theta(L_A-x),
    \end{equation}
    where $\Theta(x)$ is the Heaviside step function. This is because $e^{i\alpha \hat{n}_i}\hat{b}_i=e^{-i\alpha}\hat{b}_ie^{i\alpha \hat{n}_i}$ and the identification of $b\sim e^{-i\theta}$. Therefore, the computation of the FCS is mapped to evaluating the path integral with the quadratic action Eq. \eqref{eq:eff} under the boundary condition Eq. \eqref{eq:bdy}. For $\alpha=0$, vortex pairs are confined in the superfluid phase, and the dominant contribution contains no vortices at long distances \cite{altland_condensed_2023}. However, a finite $\alpha$ imposes a non-trivial winding of $\theta(x,\tau)$. As an example, by integrating $\nabla \theta$ along the yellow loop enclosing $(0,0)$ in FIG. \ref{fig2}, we find
    \begin{equation}
    W=\oint_C d\bm{l}\cdot \bm{\nabla} \theta = -\alpha +2\pi n , \ \ \ \ \ \ n\in \mathds{Z}.
    \end{equation}
    Therefore, a vortex exists at $(0,0)$. Similarly, we expect the presence of an anti-vortex at $(L_A,0)$ with a winding number of $-W$. For each configuration of $\theta(x,\tau)$ with fixed $n$, $\mathcal{F}_A\propto {K(2\pi n-\alpha)^2}\log L_A/{2}$ is equivalent to the increase in free energy due to the presence of these vortices \cite{zhai2021ultracold}. Summing up contributions with different $n$, we find 
    \begin{equation}\label{eq:fcscomplete}
        \mathcal{Z}_A(\alpha)=\sum_{n\in Z}Z_A^{\text{LL}}(2\pi n-\alpha)=\sum_{n\in Z}e^{-K(2\pi n-\alpha)^2\ln L_A/2+O(L_A^0)}.
    \end{equation}
    This result exhibits the $2\pi$-periodicity in $\alpha$ if we extend its domain to $\alpha \in R$, consistent with the microscopic definition.

     For sufficiently large $L_A$, the FCS is dominated by a single $n$ that minimizes the interaction energy. Therefore, away from $\alpha=\pi$, Eq. \eqref{eq:fcscomplete} is dominated by $n=0$ and reduced to $\mathcal{Z}_A^{\text{LL}}(\alpha)$. However, for $\alpha \sim \pm\pi$, two nearly degenerate configurations become dominant, as illustrated in FIG. \ref{fig2} (b-c), depicting $\theta(x,t)$ as an in-plane spin. In particular, (b) corresponds to a configuration with $n=1$ and $W=2\pi-\alpha$, while (c) corresponds to $n=0$ and $W=-\alpha$. Other terms in Eq. \eqref{eq:fcscomplete}, representing configurations where $n \neq 0, 1$, become negligible due to the large size of $L_A$, even at $\alpha=\pi$. Thus, we observe that the first-order transition between configurations (b) and (c) is the origin of the non-analytic cusp between $\pi-\epsilon$ and $-\pi+\epsilon$ as $L_A \to \infty$. With finite-$L_A$ corrections, the contributions from both terms become comparable when $|\delta\alpha|\lesssim (\ln L_A)^{-1}$, effectively smoothing out the transition at $\alpha=\pi$. This is very similar to the celebrated Page curve \cite{page_information_1993, page_time_2013}, which receives $O(1)$ corrections when the subsystem comprises exactly half of the total qubits. This analysis of the finite-size correction explains the reason for choosing the specific order of limits in our definition Eq. \eqref{eq:Delta} for extracting the non-analyticity of the FCS.

     \vspace{5pt}
    \emph{ \color{blue!60}Higher dimensions.--} 
    We then turn to higher-dimensional superfluids, where a Luttinger Liquid-type calculation is not available. In such instances, the semi-classical picture proves particularly valuable when extending our findings to higher dimensions. Here, we exemplify the case with $D=2$. Taking a finite subregion $A$, its boundary $\partial A$ forms a closed 1D loop, analogous to endpoints in 1D. Consequently, the vortex pair is replaced by a vortex loop situated at $\partial A$. An illustration for a disk-like subsystem $A$ is depicted in FIG. \ref{fig2} (d), where we assume the configuration of the phase field is independent of the azimuthal angle $\varphi$. It is established that the excitation energy for a vortex loop with a winding angle $W$ can be approximated by $W^2 L_A\ln L_A$ \cite{zhai2021ultracold}. Therefore, upon summing up all conceivable configurations characterized by different winding numbers, we deduce:
    \begin{equation}\label{eq:res2D}
        \mathcal{Z}_A(\alpha)\sim \sum_{n\in Z}e^{-C (\alpha-2\pi n)^2 L_A\ln L_A+O(L_A)},
    \end{equation}
    with a coefficient $C \propto \rho_s$. We include a potential local UV contribution, which is proportional to the boundary length $|\partial A|\sim L_A$. If we take the limit of large $L_A$ first, the result reduces to a quadratic function for $\alpha\in(-\pi,\pi)$ similar to 1D, which leads to a cusp at $\alpha=\pi$. For large but finite $L_A$, the cusp gets smoothed out with a width $|\delta\alpha|\lesssim(L_A\ln L_A)^{-1}$, which is much narrower than its counterparts in 1D. Results for more general spatial dimensions only require replacing $L_A\ln L_A$ with $L_A^{D-1}\ln L_A$, accounting for the energy of a topological excitation with (spacetime) codimension 2.

     \vspace{5pt}

    \begin{figure}
    \centering
    \includegraphics[width=0.99\linewidth]{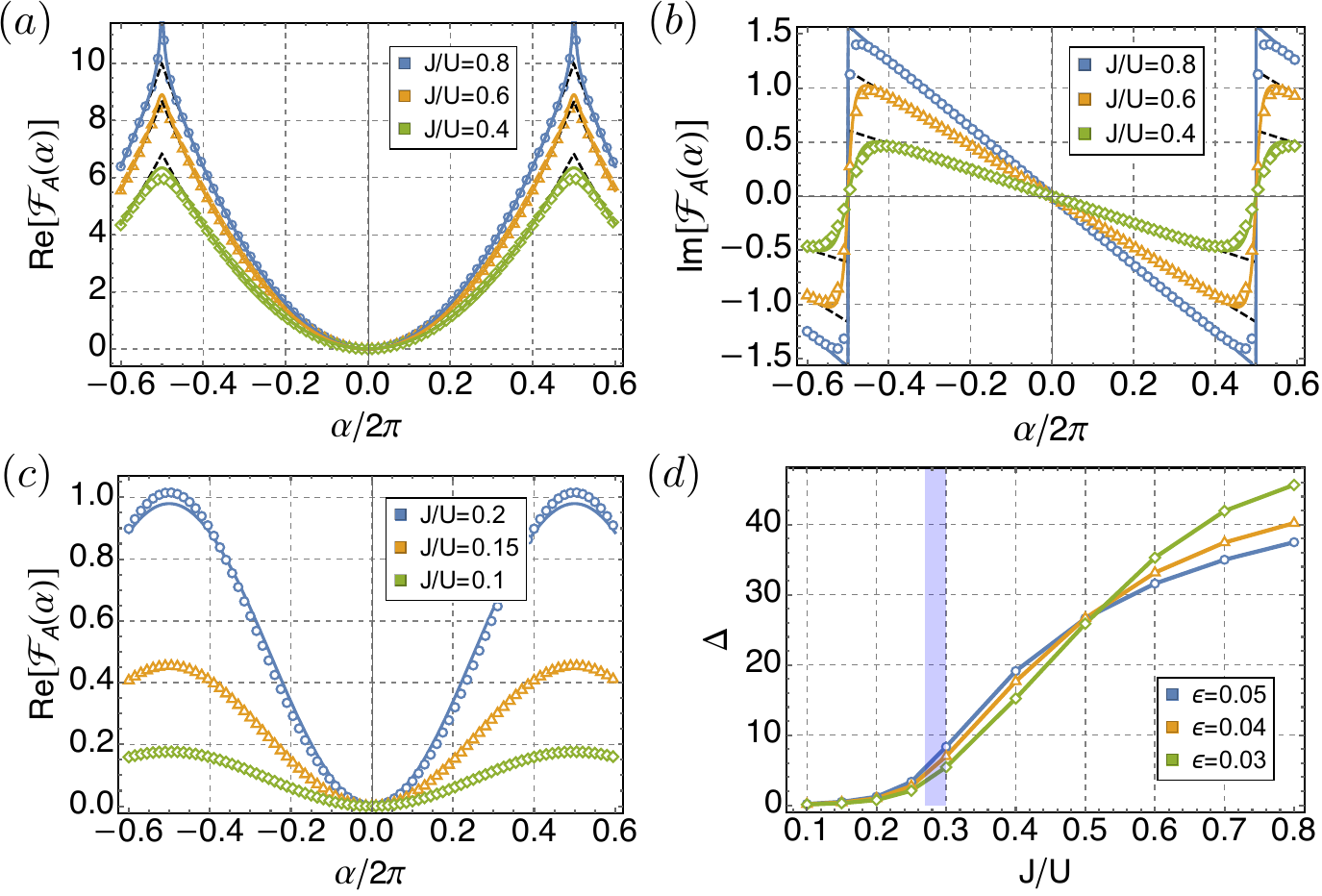}
    \caption{ Numerical results for the FCS of the Bose-Hubbard model in 1D, obtained through MPS simulation with a system size of $L=100$ in both the superfluid phase (a-b) and the Mott phase (c). In (a-b), the solid and black dashed lines represent the theoretical predictions with and without finite-size corrections, respectively, as elaborated in the main text. In (c), the solid lines represent the fitting with $C_M(1-\cos(\alpha))$. Additionally, in (d), we plot $\Delta$ as a function of $J/U$ for various small but finite $\epsilon$. The shaded region indicates the transition point reported in previous numerics \cite{PhysRevB.58.R14741}. For clarity, we periodically extend the domain of the FCS to the $\alpha \in [-1.2\pi,1.2\pi]$ to better observe FCS behavior near $\alpha = \pm \pi$.} \label{fig3}
\end{figure}
     
    \emph{ \color{blue!60}Mott phase.--}
    We now turn to compute the FCS in the Mott phase. Since the charge fluctuation is heavily suppressed, we perform a perturbative study in terms of $J/U \ll 1$. We introduce $\hat{H}_J = -\sum_{\langle ij\rangle} (\hat{b}_i^\dag \hat{b}_j + h.c.)$. In first-order perturbation theory, the ground state of the Bose-Hubbard model in arbitrary space dimension with integer filling is approximated as:
\begin{equation}\label{eq:perturbation}
|\Psi\rangle \approx \frac{1}{\mathcal{N}}\left(|\Psi_0\rangle - \frac{J}{U}\hat{H}_J |\Psi_0\rangle\right),
\end{equation}
where $|\Psi_0\rangle = \prod_i (b_i^\dag)^{\overline{n}}|0\rangle$ is the state with $\overline{n}$ particles on each site, and $\mathcal{N}$ is the normalization factor. The result is valid in arbitrary space dimensions.
By noticing that only the hopping in the boundary changes the total particle number in the region $A$, we have
\begin{equation}
\begin{aligned}
\mathcal{Z}_A(\alpha)  
&\approx 1 - \frac{4J^2}{U^2}\overline{n}(\overline{n}+1)(1 - \cos\alpha) |\partial A|,\\
\mathcal{F}_A(\alpha) & \approx \frac{4J^2}{U^2}\overline{n}(\overline{n}+1)(1-\cos\alpha) |\partial A|,
\end{aligned}
\end{equation}
This is a continuous function of $\alpha$. As a result, we determine that $\Delta=0$ holds true for $J/U \ll 1$. 

For larger $J/U$, we should compute the FCS to higher orders in $J/U$. The $m$-th order perturbation theory may excite $m$ doublons near the boundary $\partial A$, which contributes to a contribution $\mathcal{F}_A^{(m)}(\alpha)\propto (J/U)^m[1-\cos(m\alpha)]$. It is reasonable to assume the expansion converges absolutely in the Mott phase. 
Based on this argument, we anticipate that having a smooth FCS will be a generic characteristic within the Mott phase, as verified by the numerical simulations below. 
\vspace{5pt}

    \emph{ \color{blue!60}Numerics.--}
We conduct numerical simulations of the Bose-Hubbard model in both 1D and 2D using the Matrix-Product States (MPS) representation \cite{schollwock_densitymatrix_2011, ostlund_thermodynamic_1995, perez-garcia_matrix_2007}, implemented through the \texttt{ITensors.jl} package \cite{10.21468/SciPostPhysCodeb.4}. We have checked that the bond dimension is large enough to reach convergent results. In 1D, we fix the system size at $L=100$ with unit filling $\overline{n}=1$ and open boundary conditions and select subsystem $A$ such that it consists of $L_A=35$ contiguous sites at the center. The results of $\mathcal{F}_A(\alpha)$ in both phases are presented in FIG. \ref{fig3} (a-c). For clarity, we extend the plot range of $\alpha/2\pi$ slightly to $[-0.6, 0.6]$ by utilizing its periodicity. Unlike the effective theory Eq. \eqref{eq:eff} or the perturbative analysis Eq. \eqref{eq:perturbation} which exhibits an emergent particle-hole symmetry, the microscopic Hamiltonian explicitly breaks this symmetry. As a consequence, $\mathcal{F}_A(\alpha)$ acquires a non-universal imaginary part, whose magnitude is much smaller than the real part.

In the superfluid phase, $\text{Re}[\mathcal{F}_A]$ closely resembles a quadratic function of $\alpha$, as predicted by Eq. \eqref{eq:Fquadratic}. We fit $\text{Re}[\mathcal{F}_A]$ with the functional form $C_R\alpha^2$ for $\alpha/2\pi \in [-0.4,0.4]$. The result is depicted as a dashed black line in FIG. \ref{fig3} (b), which exhibits a high degree of accuracy in matching the numerics away from $\alpha=\pm \pi$. Within the same range of $\alpha$, we observe that $\text{Im}[\mathcal{F}_A]= C_I\alpha$, as indicated by the black dashed lines in FIG. \ref{fig3} (b). In Eq. \eqref{eq:fcscomplete}, the imaginary part of $\mathcal{F}_A$ is encompassed within the non-universal contribution $O(L_A^0)$ in Eq. \eqref{eq:fcscomplete}. Although it is an order of magnitude smaller than the real part $\text{Re}[\mathcal{F}_A]$, it still introduces significant finite-size corrections when compared to the numerics. We propose the following expression for $Z_A(\alpha)$:
\begin{equation}
    Z_A(\alpha) = \sum_{n \in \mathbb{Z}} e^{-C_R(\alpha - 2\pi n)^2 - iC_I(\alpha - 2\pi n)},
\end{equation}
which is plotted in both (a) and (b) as solid lines, demonstrating good accuracy even near $\alpha = \pm\pi$. This indicates the presence of non-analyticity in the large $L_A$ limit. For comparison, the FCS in the Mott phase can be approximated by $C_M(1-\cos \alpha)$, which is a continuous function near $\alpha=\pm \pi$. This is shown in FIG \ref{fig3} (c). Moreover, $\text{Im}[\mathcal{F}_A]$ is smaller than $10^{-1}$ throughout the Mott phase. Finally, we determine $\Delta$ for different $J/U$ by taking numerical derivatives at small but finite $\epsilon$. Our proposal for using $\Delta$ as a non-local order parameter is supported by the results in FIG \ref{fig3} (d).

\begin{figure}
    \centering
    \includegraphics[width=0.99\linewidth]{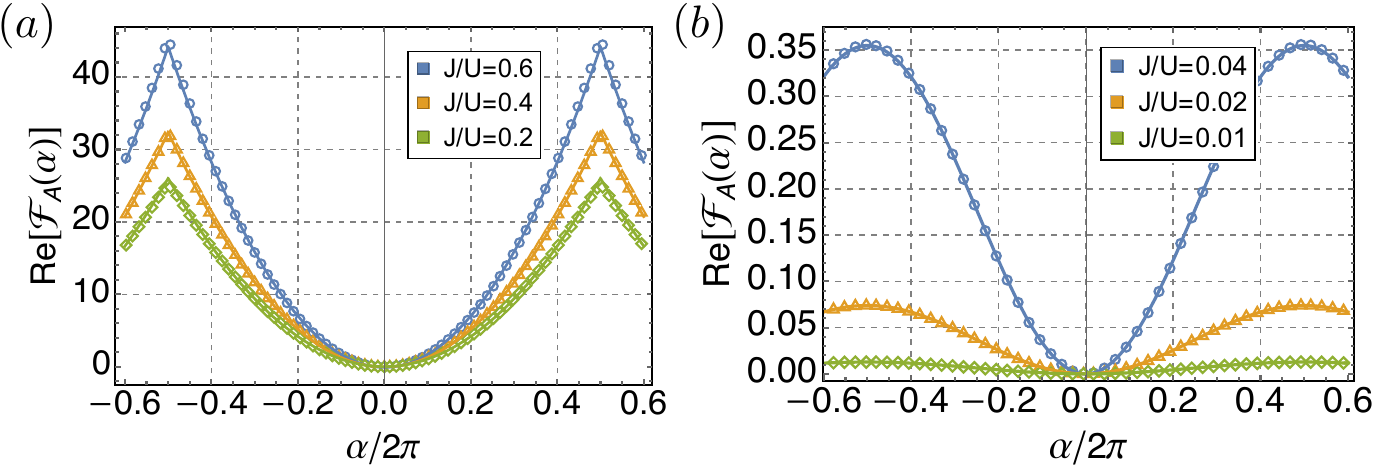}
    \caption{ Numerical results for the FCS of the Bose-Hubbard model in 2D, obtained through MPS simulation with a system size of $L_x=25$ and $L_y=6$ in both the superfluid phase (a) and the Mott phase (b). In (a), the solid lines represent the theoretical predictions without finite-size corrections, which already show excellent agreement with the numerical data. In (c), the solid lines represent the fitting with $C_M(1-\cos(\alpha))$. For clarity, we periodically extend the domain of the FCS to the $\alpha \in [-1.2\pi,1.2\pi]$ to better observe FCS behavior near $\alpha = \pm \pi$.} \label{fig4}
\end{figure}

To further test our proposal in higher dimensions, we investigate the FCS in the 2D Bose-Hubbard model on a strip with $L_x=25$ and $L_y=6$, with an open boundary condition in the $x$ direction and a periodic boundary condition in the $y$ direction. The subsystem $A$ is a rectangular region in the center of the system with $L_{xA}=6$ and $L_{yA}=3$. Previous studies report the superfluid-to-Mott transition occurs at $r_c\approx 0.06$ \cite{capogrosso-sansone_monte_2008, chen_universal_2014, soyler_phase_2011}. The numerical results are presented in FIG. \ref{fig4}. In the superfluid phase, solid lines correspond to theoretical predictions without finite-size corrections $\text{Im}[\mathcal{F}_A]\propto \alpha^2$ for $\alpha\in(-\pi,\pi)$. Despite a small subsystem size, we find that numerical results match our theory with good accuracy. This demonstrates a parametrically smaller finite-size broadening compared to the 1D case in FIG. \ref{fig3} (a), consistent with Eq. \eqref{eq:res2D}. We also present results in the Mott phase in FIG. \ref{fig4} (b), which can be well approximated by $C_M(1-\cos \alpha)$

\vspace{5pt}

    \emph{ \color{blue!60}Discussions.--}
    In this letter, we investigate the FCS of U(1) conservation charges in systems undergoing quantum phase transitions, using the Bose-Hubbard model as a concrete example. By employing an effective theory description, we demonstrate that the FCS exhibits a cusp near $\alpha=\pm \pi$  in the superfluid phase. This cusp originates from a first-order transition between distinct vortex configurations in the ordered phase. As a result, the discontinuity in the first-order derivative acts as a non-local order parameter for the superfluid-to-Mott transition. Our theoretical proposal is supported by Matrix Product State (MPS) simulations in both one-dimensional (1D) and two-dimensional (2D) cases and can be readily verified in state-of-the-art experiments \cite{PhysRevX.13.021042}.

    To further strengthen our proposal, it would be beneficial to conduct a Monte Carlo simulation in higher dimensions with larger system sizes, a task we plan to undertake in future studies. While we have focused on the superfluid-to-Mott transition as an illustrative example, vortices naturally arise in all systems exhibiting U(1) symmetry breaking. Consequently, we anticipate the emergence of cusps in the ordered phase to be a general characteristic in such scenarios, at least for models with on-site U(1) symmetries. It would be intriguing to investigate whether this phenomenon can be generalized to models with non-Abelian symmetry groups, such as SU(2) symmetry, or to systems with generalized symmetries.

\vspace{5pt}
    \emph{Acknowledgement.} 
    We thank Zhen Bi, Meng Cheng, Yingfei Gu, Shenghan Jiang, Liang Mao, and Hui Zhai for helpful discussions. We especially thank Hui Zhai for providing invaluable suggestions to improve the manuscript. This project is supported by NSFC under Grant No. 12374477.

\vspace{5pt}
    \emph{Note added.} 
    While finalizing our manuscript, we became aware of related investigations on FCS \cite{newpaper1,newpaper2,newpaper3}.

\bibliography{draft.bbl}
    
\end{document}